\documentstyle[preprint,aps]{revtex}
\begin{document}
\draft
\title{Calculated Schwoebel barriers on Si(111) steps using an empirical 
potential}
\author{S.\ Kodiyalam, K.E.\ Khor and S.\ Das Sarma}
\address{Department of Physics, University of Maryland, College Park, Maryland 
20742-4111}
\address{\rm (Submitted to Physical Review on July 14th 1995, Resubmitted on 
November 27 1995)}
\address{\mbox{ }}
\address{\parbox{16cm}{\rm \mbox{ }\mbox{ }\mbox{ }
Motivated by the recent investigations on  
instabilities caused by Schwoebel barriers during growth and their effects 
on growth or sublimation by step flows, we have investigated,
using the Stillinger-Weber potential, how this step edge barrier arises for 
the two high 
symmetry steps on 1$\times$1 reconstructed Si(111). Relative to a barrier of
0.97 $\pm$ 0.07 eV on the surface, we find additional (Schwoebel) barriers of 
0.61 $\pm$ 0.07 eV and 0.16 $\pm$ 0.07 eV for adatom migration over the 
[$\overline{2}11$]  and the [$\overline{1}\overline{1}2$] steps respectively.
The adatom potential energy is found to be strongly correlated with that
derived from the local 
geometry of atoms on the adatom-free surface or step edges. This correlation 
preserves a strict correspondence between the barrier determining features in 
the spatial variation
of the adatom potential energy and the same derived from the local geometry 
for the Si(111) surface and the [$\overline{2}11$] step. It is therefore argued
that the Schwoebel barrier on the [$\overline{2}11$] step is robust {\it i.e.}
a feature that would survive in  more satisfactory {\it ab initio} or tight 
binding calculations. Using a diffusion equation for the adatom 
concentration the relevance of 
the barrier to electromigration of steps has been explored. Data from such 
experiments on Si(111) has been used to place an upper bound on the 
Schwoebel barrier and a lower bound on the electromigration force.}}
\address{\mbox{ }}
\address{\mbox{ }}
\address{\parbox{16cm}{\rm PACS numbers: 68.35Fx, 68.35Ja, 68.35Bs, 68.35Md }}
\maketitle


\bigskip
\narrowtext

\section{Introduction}
             
The Schwoebel barrier was originally introduced in the context of 
step motion \cite{Shsp} as the additional barrier for adatom diffusion 
over a step edge from the upper to lower terraces. It was argued that such a
barrier results in an anisotropy in adatom diffusion into the step edge - the 
diffusion from the lower terrace being greater. This anisotropy was found to 
drive an arbitrary distribution of 
step spacings towards a uniform distribution during growth of a vicinal 
surface.\cite{Shsp}
Later, it was pointed out by Villain\cite{Vill} that this 
growth by step flow is stable only on a sufficiently vicinal surface with 
possible instabilities setting in during the growth of a flat (singular) 
surface. The dynamical morphology of a singular surface growing under the 
influence of a schwoebel barrier is a subject of great current activity. 
\cite{Mdj,MsMp,Awh,Ie,Cjl}  While the eventual fate of growth on a flat 
surface in the presence of such extra step edge barriers is still being 
discussed \cite{Mdj,MsMp,Awh,Ie,Cjl}, it is now well accepted that 
Schwoebel barriers lead to coarsening in the evolving surface morphology 
under nonequilibrium growth conditions, producing mounds, pyramids and 
facet-like angular structures on the growing surface. Recent experimental 
studies of nonequilibrium growth on the Si(111) surface have produced 
somewhat contradictory \cite{Jc,Hny,Cjl1} results, and the specific role of 
Schwoebel barriers for nonequilibrium Si(111) growth is unclear at the 
present time.

The current study is primarily motivated by observations of another 
instability: As was pointed out by Schwoebel and Shipsey \cite{Shsp}, an 
anisotropy favoring diffusion into the step edge from the upper terrace 
(possibly due to larger barriers for diffusion from the lower terrace) results 
in a step pairing instability during the growth of a vicinal surface.
In recent experiments similar direct-current induced
reversible step-bunching instabilities have been observed during sublimation of
the high temperature 1$\times$1 phase of Si(111).\cite{Avl,EFu}. Here, we 
report on a calculation of 
the Schwoebel barrier for the two high symmetry vicinal steps on 1$\times$1 
Si(111) using the empirical Stillinger-Weber potential.\cite{Stil} The 
importance of the Schwoebel barrier in the context of the electromigration 
experiments has also been explored by modeling the barrier as affecting the 
boundary conditions  to the diffusion equation for adatom concentration 
used to interpret these experiments \cite {Sss}.

\section{Empirical potential calculations}

The use of the Stillinger-Weber potential in this study has been motivated 
by its successful application in previous studies of bulk and liquid 
silicon \cite{Stil}, the Si(100) surface and steps on this surface \cite{Tw}.
Barriers on the Si(100) surface and on single and double height steps on this 
surface have also been calculated using this potential. \cite{RG1,RG2,RG3} 
Although this potential fails to reproduce the correct energetics of the 
Si(111) surface configurations with adatoms, \cite{Li} it has been used here 
since features that follow purely from the changes in the adatom coordination 
number are expected to be robust i.e. these features survive even if the 
details of the empirical potential used change. Such features are 
expected to survive in more satisfactory {\it ab initio} or tight binding 
calculations. The calculation here is followed by an attempt to 
identify such robust features. 
  
To determine the diffusion barriers the adatom potential energy has been mapped 
as a function of the ($x,y$) position of the adatom (in the (111) plane) 
for the Si(111) surface (Fig. 2(a)), the [$\overline{2}11$] step (Fig. 2(d))
and the [$\overline{1}\overline{1}2$] step (Fig. 2(g)). The
three-fold and reflection symmetry of the Si(111) as shown 
in Fig. 1 implies that steps running along directions with equal $\theta$ 
are identical. It has been 
shown in a previous study \cite{Skod} using the Stillinger-Weber potential 
that an alternative configuration of 
step-edge atoms with some of the upper terrace atoms rebonding to atoms in the 
lower terrace gives a lower step energy for the [$\overline{1}\overline{1}2$]
and [$\overline{1}01$] steps. However, such a configuration has not been 
considered here
since it has also been shown that it gives rise to  
step-step interactions an order of magnitude larger than experimental 
estimates.\cite{Skod} 
Neglecting such rebonded configurations, all the intermediate low symmetry 
steps ($0^0 < \theta < 60^0$) have been shown to have a higher step energy.
\cite{Skod} In other words, diffusion barriers have been 
calculated for those steps whose interactions are not larger than experimental 
estimates and whose step energy is a local minimum  as a function their 
orientation $\theta$. The adatom potential energy $V$ has been computed as the 
difference in the minimum potential energy of the system with the adatom at 
infinity (non-interacting) and the same with the interacting adatom.  

Standard molecular dynamics (MD) procedures of integrating Newton's law (with 
dissipation to reduce temperature) and  the steepest descent equations have 
been used to determine the minimum potential energy of the system. \cite{Skod} 
These routines determined the adatom potential energy to an accuracy of 
$10^{-4}$ eV. The ($x,y$) 
coordinates of the adatom are fixed during the integration process.
The system consisted of a certain number of bi-layers of Si(111) in an MD cell 
with the surface lattice constants $a_1$ and $a_2$ along its $x$ and $y$ 
directions respectively (see Fig. 1). Three of the bottom bi-layers are fixed 
at bulk lattice coordinates throughout the calculation. In simulations on the 
Si(111) surface with six additional (movable) bi-layers it was found that 
changing the adatom's ($x,y$) position from a deep minimum lead to the 
shearing of the entire surface in the $xy$ plane towards the adatom. As this 
made the ($x,y$) position of the adatom relative to the surface ill defined 
further
simulations on the surface as well as the step configurations were carried out 
with atoms at the ($x,y$) boundaries also fixed at positions corresponding to 
the adatom-free but relaxed configurations. 
The regions in the (111) plane explored in all cases were ``centrally'' 
located i.e.
maximally away from the boundaries so that finite size effects are minimized.
System size dependence in $V(x,y)$ was initially 
explored for the Si(111) surface to determine the optimal system size.
It was found that changing the system size from four  lattice constants in the 
$x$ and $y$ directions (including the boundary of fixed atoms) with three 
movable bi-layers to six  lattice constants in the $x$ and $y$ directions 
with six movable 
bi-layers changed (reduced) the adatom potential energy (at the minima, maxima 
and saddle points) by $<10^{-2}$ eV. As an error bar of $\pm 10^{-2}$ eV
was estimated to be smaller that the accuracy needed for this study all the 
simulations were carried out with the smaller system size - the size in 
the $y$ direction being extended to $4\frac{2}{3}$ and $4\frac{1}{3}$ lattice
constants for the [$\overline{2}11$] and [$\overline{1}\overline{1}2$] step 
configurations respectively so that periodic boundary conditions could 
be applied to create vicinal steps. Since the bulk terminated Si(111) surface
and [$\overline{2}11$] step configurations do not relax under the 
Stillinger-Weber potential, with the system sizes chosen here their 
corresponding surface and step energies are reproduced exactly. However, the 
[$\overline{1}\overline{1}2$] step configuration is different from the 
bulk terminated structure due to the 2$\times$1 reconstruction at the step 
edge. With the system size chosen here it's 
step energy is reproduced to within 4$\times10^{-4}$ eV/$a_1$.  \cite{Skod}

The MD procedures began with initial configurations 
for each ($x,y$) position of the adatom corresponding
to the relaxed adatom-free structures.
The $z$ coordinate of the adatom was chosen to be equal to that obtained 
in the final configuration during simulation with the adatom in the 
immediate neighborhood of the point ($x,y$). For the step configurations 
$V(x,y)$ was determined from behind to the front of the step edge.
These procedures for determining $V(x,y)$ were found 
necessary for the [$\overline{2}11$] step configuration since other methods 
such as an arbitrary initial $z$ coordinate for the adatom or determining 
$V(x,y)$ from the front to behind the step edge lead to the adatom relaxing 
into configurations in which it displaces an atom near the step edge and/or 
moves into the bulk.  
Symmetries in the (111) plane were exploited to reduce the size of the regions 
in which the adatom potential energy needs to be computed. Therefore for 
the Si(111) 
surface only a sixth of the surface unit cell was explored. For this case 
$V(x,y)$ was computed on a triangular grid - the sides of the triangle 
coinciding with the high symmetry directions with the distance between 
neighboring grid points being $\frac{a_2}{9}$. An interpolation scheme 
respecting the symmetries on the surface was used to determine the features 
of $V(x,y)$. For the [$\overline{2}11$] and [$\overline{1}\overline{1}2$] step 
configurations reflection symmetry about the $y$ axis ($\perp$ to the step edge)
reduced the width of the region (along the step edge) explored to a length 
of $\frac{a_1}{2}$ and $a_1$ respectively. For these configurations $V(x,y)$ 
was determined on  
a rectangular grid with the distance between neighboring grid points being 
$\frac{a_1}{16}$ and $\frac{a_2}{30}$ along the $x$ and $y$ axis respectively.
With the step edge in the middle, the length of the region explored was 
$1\frac{2}{3}a_2$ and $1\frac{1}{3}a_2$ for the [$\overline{2}11$] and 
[$\overline{1}\overline{1}2$] steps respectively. This was to enable a new 
interpolation scheme (respecting the symmetries of the step 
configurations) applying periodic boundary conditions along the $x$ axis
to do the same along the $y$ axis of the region explored. On comparing the 
potential energy of the adatom on the Si(111) surface to that far away from the 
[$\overline{2}11$] step edge errors due to the finite grid sizes, 
the interpolation schemes and possibly effects due to being close to the 
boundary of fixed atoms were recognized. 
These errors are 
conservatively estimated to be $\pm$0.05 eV. Since this is much larger than 
the errors due to finite size effects it is assumed to be the error bar in the 
potential energy. Barrier values being differences in these potential 
energies are 
therefore estimated to have an error bar of $\pm$0.07 eV.

\section{Results and discussion}

The results are shown in Fig. 2. The global minimum of the adatom potential
energy $V$ occurs on the $H_3$ site (m$_1$) on the Si(111) surface (Fig. 2(b)) 
where the adatom potential energy is
-3.31 eV. This is a significant fraction of the bulk energy per atom: -4.34 eV.
The relevant saddle point for $H_3 \leftrightarrow H_3$ transitions is s$_1$
(close to the $T_4$ cite which is a local minimum)
where the





potential energy is -2.34 eV. Thus the barrier to diffusion on the
surface is 0.97$\pm$0.07 eV. These results are consistent
with previous studies (using the same potential)
on surface energies of configurations with adatoms \cite{Li} as well as a
study of diffusion on the Si(111) surface. \cite{KD} The barriers to diffusion
between the minima along the step edges are slightly smaller than the
barrier on the surface - apparently inconsistent with previous work \cite{NC}
showing that step edge fluctuations are (predominantly) due to
attachment/detachment of adatoms from the terrace and not due to diffusion
of atoms along the step edge.
In this study the
Schwoebel barrier has been defined to be difference between the maximum
adatom potential energy (along the path on which this is a minimum) as it moves
into the step edge from the global minimum on the upper terrace far away from
the step edge and the same for the $H_3 \leftrightarrow H_3$ transitions on
the Si(111) surface. In other words it is the difference in the adatom
potential energies at the barrier determining saddle point on the step
configurations and the same on the free surface (s$_1$). With this definition
the Schwoebel barrier cannot be negative. From Fig. 2(e) (Fig. 2(h)), for  the
[$\overline{2}11$] ([$\overline{1}\overline{1}2$]) step the barrier is
determined by s$_3$  (s$_5$) where the adatom potential energy is -1.73 eV
(-2.18 eV) - implying a Schwoebel barrier of 0.61$\pm$0.07 eV
(0.16$\pm$0.07 eV). Growth on Si(111) is therefore expected to produce 
moundlike structures with facets consisting (predominantly) of 
[$\overline{2}11$] steps. 
However, the experiments of Yang {\it et al} (temperature = 275$\pm$5$^0$ C) 
instead show facets with 
[$\overline{11}2$] steps \cite{Hny}. This discrepancy may be due to 
the presence of the 7$\times$7 reconstruction.

It must be noted that the Stillinger-Weber potential has been tuned 
only to the properties of the bulk diamond and liquid structures of Silicon 
and not to any surface or step properties. As mentioned previously it does 
not produce the correct energetics for the Si(111) surface configurations with 
adatoms \cite{Li} as compared to {\it ab initio} calculations. 
\cite{Nor,Nor1,KD2} Therefore an attempt has been made to identify the robust 
features of 
this study: Features following from changes in the coordination number
of the adatom. This idea is supported by the observation that the 
reconstruction energy of the Si(100) surface calculated using the 
Stillinger-Weber potential \cite{KD3} (0.85 eV) agrees with {\it ab initio} 
calculations \cite{YC} (0.84 eV). The Tersoff and Dodson empirical 
potentials also give this energy to be of the same order of magnitude. 
\cite{KD3} The Stillinger-Weber also reproduces the correct order of energies 
of the 
[$\overline{2}11$] and [$\overline{1}\overline{1}2$] steps (per step edge atom 
these values are 0.72 eV and 0.62 eV respectively\cite{Skod}) as well as the 
presence of rebonding at the [$\overline{1}\overline{1}2$] step edge as 
compared to tight binding calculations (per step edge atom these values are 
0.70 eV and 0.38 eV respectively \cite{Cha}). During reconstruction, the 
coordination number of atoms (on the (100) surface and 
[$\overline{1}\overline{1}2$] step edge) changes from 
two to three. It is therefore expected that features following from 
such a change in coordination number are not artifacts of 
the empirical potential used. Here, the adatom energy recomputed 
$without$ $additional$ relaxation of other atoms due to the presence of
the adatom is assumed to be a good measure of the coordination number. 
Although this measure is very similar to the actual adatom energy $V(x,y)$, it 
helps to identify features that follow from the geometry of atoms (locally 
around to adatom) on the relaxed adatom-free surface or step edges. Features 
that do not change significantly due to additional relaxations in the presence 
of the adatom are expected to be robust. The adatom 
potential energy $V_{lg}(x,y)$ has therefore 
been recomputed 
with other atoms fixed at positions corresponding to the relaxed adatom-free 
structures. These results are shown in Fig. 2(c) for the Si(111) surface, 
Fig. 2(f) for the [$\overline{2}11$] step and in Fig. 2(i) for the 
[$\overline{1}\overline{1}2$] step. Similarities in the contour plots of 
$V$ and $V_{lg}$ suggest a strong correlation between them. This correlation 
is evident from the linear relationship between $V_{lg}$ and $V$ (Fig. 3)
The lines in Fig. 3 are best fits to $V_{lg}(x,y)$ $vs$ $V(x,y)$ which were
explicitly computed 
at the grid points. There is no repetition of points that are equivalent 
due to the symmetries of the relevant configurations. Further, for the step 
configurations, the length of the region along the $y$ axis corresponding 
to these plots is $\frac{1}{2}a_2$ on either side of the step edge. The 
rough configuration independence of the relationship between $V_{lg}$ and $V$ 
suggests that it is a characteristic of the [111] surface. A similar 
behavior may be true for the Si(100) surface and single height steps on it.   
In this case, borrowing the values of $V$ corresponding to the features 
(minima and saddle points) from previous work by Roland and Gilmer 
\cite{RG1,RG2,RG3}, and computing $V_{lg}$ for the same features here, it is 
found that $V_{lg}=(1.3\pm 0.2)V+(1.2\pm 0.5)$ for the Si(100) surface and 
 $V_{lg}=(1.1\pm 0.1)V+(0.6\pm 0.3)$ for the combined data from the three 
single height steps.

\subsection{Relevance to electromigration experiments}

Recent experiments on the electromigration of steps on Si(111)\cite{Avl,EFu} 
can be reinterpreted in the presence of a Schwoebel barrier by modifying the 
boundary conditions to the diffusion equation for adatom concentration 
 used previously by Stoyanov {\it et al} \cite{Sss} to describe step bunching
 (see appendix A). 
The particular modification is to suppress the strength of the step as a 
source of adatoms onto the upper terrace relative to the same onto the lower 
terrace by a factor $e^{\epsilon}$ where $\epsilon = \frac{2E_s}{k_bT}$ with 
$E_s$ being the Schwoebel barrier. With this modification the 
equations for adatom concentration on a particular terrace has four length 
scales - the diffusion length $\lambda$ and the scale 
introduced by the electromigration force $f$ (=$\frac{F}{k_bT}$)  both of 
which are parameters entering the diffusion equation and 
two other scales $t_0$ and $t_1$ entering the boundary condition to this 
equation at the upper and lower terrace step edges respectively 
($t_{0,1}=\frac{\beta_{0,1}}{a^2Dn_e}$, $a^{-2}$ is the density of atoms in a 
terrace, $D$ is the diffusion constant, $n_e$ is the equilibrium adatom density 
and $\beta$ is the step kinetic coefficient - here the 
Schwoebel barrier is modeled as making this parameter assume different values 
at the upper and lower terrace step edges). 
These are in addition to the scale introduced  by the terrace width $W$.
Since the modification is further designed to keep the total strength of the 
step as a source of adatoms a constant independent of $E_s$, $t_0+t_1=t$ - a 
constant independent of $\epsilon$ for fixed $k_bT$ (variation of $t$ with 
$E_s$ is included in the estimated range of $t$ ) with $\frac{t_0}{t_1}=
e^\epsilon$. The (linear) diffusion equation is then solved, with
$n_e$ appearing in the boundary conditions as a 
``source'', to obtain the spatial variation of the adatom concentration.

Using the above solution the equations for the velocities 
of an array of steps with positions $X_i$ are developed- the step index $i$ 
increases in the step down direction which is also the positive $x$ axis. 
The time scale that enters these equations is the lifetime of 
an adatom $\tau_{lf}$. These equations have the following form (see appendix A):

\begin{equation}
\dot{X}_n = \frac{\theta_e}{\tau_{lf}}\lambda[g_-(\epsilon,t\lambda,f\lambda,
W_n\lambda^{-1})
+ g_+(\epsilon,t\lambda,f\lambda,W_{n-1}\lambda^{-1})] \
\end{equation}

\noindent where $\theta_e=n_ea^2$ and the terrace 
width $W_n=X_{n+1}-X_n$.
From the above, equations for the rate of change of the terrace widths can be 
obtained. Linear stability analysis of these equations around an average 
terrace width $w$ would predict when the step bunching instability occurs. The 
linearized equations may be written as follows:

\begin{equation}
\dot{\delta}_n = k_-\dot{\delta}_{n+1} + (k_+ - k_-)\dot{\delta}_n + k_+\dot{
\delta}_{n-1} \ 
\end{equation}

\noindent where $\delta_n=W_n-w$ and 
$k_\pm=\frac{\theta_e}{\tau_{lf}}\lambda[\frac{\partial g_\pm}{\partial W}]_
{W=w}$. The instability occurs if \cite{BG} $k_+ - k_- > 0$ or equivalently 
(since $k_+ + k_- <0$ for $w>0$), when the anisotropy ratio $\rho = 
\frac{k_+}{k_-} < 1$. This analysis is similar to the work of Ghez {\it et al}
\cite{G} which includes an external flux with the electromigration force 
being absent.

The recent experiment of Williams {\it et al} on $reversible$ step bunching on 
Si(111), which measures an ``effective'' anisotropy ratio $\rho_{eff}$ 
(near bunched steps) in the temperature range 1155-1215$^o$ C (this range
includes the corrections due to the emissivity of the optical pyrometer 
used in the experiment \cite{EFu}), can now be used to  
to determine an upper bound on $E_s$ ($E_s^u$) and a lower bound 
on $F$ ($F_\ell$) by solving 
the equation $\tau^{-1}=k_+-\rho k_-=0$. The solution to this equation is 
obtained by estimating the parameters $w$, $t$ and $\lambda$ (in the 
temperature range of interest: 1155-1215$^o$ C) and then 
determining $f$ as a function of $\epsilon$.
The estimates of the upper and lower bounds on $t$ and $\lambda$ are made
$conservatively$ - the range is made wide enough so as to include the ``true 
value''.
Experimental observations of Latyshev {\it et al} \cite{Avl} show that
around a
temperature of 1200$^o$ C the step 
velocity varies linearly with the terrace width upto a width of 2 $\mu m$ .
A lower bound on the diffusion length \cite{E} of 0.5 $\mu m$ is obtained by 
reproducing this result using equation $1$. This estimate is robust against 
large changes in the parameters $t$ and $f$. To make this estimate, $\epsilon$ 
is conservatively chosen to be zero.  Further, to first order in 
$w\lambda^{-1}$, $g_++g_-=-w\lambda^{-1}$ - independent of $t$, $\epsilon$ and 
$f$. Therefore the magnitude of the slopes of the above curves correspond to 
the evaporation rates $r=\frac{\theta_e}{\tau_{lf}}$. An estimate of the upper 
bound on $\lambda$ is obtained by estimating the diffusion constant \cite{E} 
$D$ (using $D=b^2\nu e^{-E_a/k_bT}$, $b$=3.84$\AA$, $\nu$=10$^{13}$s$^{-1}$, 
$E_a$=0.97 eV (calculated here)) 
and an estimate of the upper bound on $\tau_{lf}=\frac{n_ea^2}{r}$ (The 
temperature dependence of $r$ in the data from Latyshev {\it et al} \cite{Avl}
is consistent with the theory of Burton {\it et al} \cite{BCF} (BCF) with an 
activation energy equal to the cohesive energy of silicon ($E_b$=4.34 eV)). 
The upper bound on $\theta_e$ is $assumed$ to be 0.167 - the primary motivation 
for this is that it is measured to be $\approx$ 0.1 around 900$^o$ C and 
\cite{YY} its activation energy $E_n$, which is equal to $E_b-E_\tau$ 
($E_\tau$ is 
the activation energy for $\tau_{lf}^{-1}$) from the BCF theory, is estimated 
to be negative since total energy calculations \cite{MV} predict smaller 
surface energies for silicon surfaces with adatoms as compared to the 
1$\times$1 surface thereby making $E_\tau>E_b$. The specific value of 0.167 
is chosen as 
it corresponds to the density in any $\surd$3$\times\surd$3
configuration of adatoms wherein all the floating bonds on the 1$\times$1 
substrate are 
saturated by bonding with the adatoms - at higher densities adatom interactions
 are $expected$ to be 
significant. This bound is therefore required here as a measure of internal 
consistency in the analysis since the diffusion equations used correspond to 
free adatoms. The upper bound on $\tau_{lf}$ 
is calculated using this value of $\theta_e$. As the variation in $n_e$ has been
argued to be small \cite{E} $E_n\approx 0$ and $E_\tau \approx E_b > E_a$. 
The estimate of the upper bound on $\lambda$ (=$(D\tau_{lf})^{1/2}$) is 
therefore made at the lowest temperature of interest and is equal to 
$\approx$ 70 $\mu m$. The upper bound used in this study is $\approx$7 times 
this value, making 0.5 $\mu m$  $\leq$ $\lambda$ $\leq$ 0.5$\times$10$^3$ 
$\mu m$.

The estimate of the range of $t$ is obtained from the range of $\lambda$, the 
measured $r(T_2)$ \cite{Avl} and $\theta_e(T_1)$ \cite{YY} and the relation 
between $\beta_s=\beta_0+\beta_1$ and the measured step mobility 
$\Gamma(T_1)$ \cite{NC} ($T_1$ corresponds to 900$^o$ C and $T_2$ corresponds to
temperatures of interest).  At the same temperature, $\beta_s=\frac{\Gamma}
{a^2}$ (see appendix B). 
From a model of attachment-detachment at the step edges (see appendix B) the 
temperature and $\theta_e$ dependences of $\beta_s$ is determined: 
$\beta_s \propto \frac{ (e^{\frac{-2E_a}{k_bT}}+e^{\frac{-2(E_a+E_s)}{k_bT}})
\theta_e^2 }{1-\theta_e}$. 
The upper bound on $\beta_s$ is derived using the 
above dependence from the lower bound = $\frac{\Gamma(T_1)}{a^2}$ 
($E_a$=0.97eV,
 $E_s$ is set equal to a value that gives the largest increase). The parameter 
$t$ is then evaluated using $t=\frac{\beta_s}{a^2Dn_e}$.
This gives upper and lower bounds on $t$: 
$t_{u,l}=\frac{\Gamma(T_1)}{a^2\lambda^2r(T_2^+)} \left( 1,\frac{r(T_2^+)
(1-\theta_e(T_1)}{r(T_2^-)(1-\theta_e(T_2^+))}
\left[ \frac{e^{\frac{-E_a}{k_bT_2^+}}T_2^+\theta_e(T_2^+)}
      {e^{\frac{-E_a}{k_bT_1}}T_1\theta_e(T_1)} \right]^2 \right) $ - where
$T_2^+$ ($T_2^-$) corresponds to the highest (lowest) temperature in the range 
of interest. The values of $r(T_2^+)$ and $r(T_2^-)$ are obtained from the 
data of Latyshev {\it et al} \cite{Avl} using $E_b$=4.34 eV, $\Gamma(T_1)$ 
from a previous measurement of Bartelt {\it et al} \cite{NC}, $E_a$ = 0.97 eV, 
$\theta_e(T_1)$=0.1 (measured by Yang {\it et al} \cite{YY}) and 
$\theta(T_2^+)$=0.167 (as assumed previously).  
This results in $\frac{7.7\times10^3\mu m}{\lambda^2} \leq t \leq 
\frac{7.7\times10^3\mu m}{\lambda^2}\times10^{3.1}$. The range of $t$ therefore 
depends on the value of $\lambda$.

The value of the parameter $w$ used here is equal to the average terrace width 
in the experiment of Williams {\it et al} \cite{EFu} (=0.15 $\mu m$) 
who determine 
$\rho_{eff}$ to be 0.20$\pm$0.03. It must be noted that the extraction 
of $\rho$ from the experiment had used a theory \cite{KW} which 
makes $\rho_{eff}= \rho(w\lambda^{-1}=0)$ ($w$ was much smaller 
than the average 
terrace width since in the theory used it corresponded to the distance 
between bunched steps). In this study it is assumed that $\rho_{eff}=
\rho(w\lambda^{-1}\neq 0)$ - this may be a $reasonable$ approximation since 
here $\lambda_{min}=$0.5$\mu m$ and the $W_{max}$ in the experimental data 
used to determine $\rho_{eff}$ was as large as 0.6 $\mu m$. This approximation 
is motivated by the need to study the solution of $\tau^{-1}=0$ including 
the full non-linearity in $w\lambda^{-1}$.

The procedure for determining $E_s^u$ and $F_\ell$ assumes that an 
electromigration force $F$ causes the step bunching instability ($\rho$=0.2) 
with $-F$ restoring the stability ($\rho \geq 1$). This assumption is 
motivated by the observation \cite{EFu} that a current in 
the step up direction causes the step bunching whereas an equal magnitude 
in the step down direction results in uniformly spaced steps. $E_s^u$ \& 
$F_\ell$ are determined by studying the solutions to $\tau^{-1}=0$ (fixing 
$t$, $\lambda$ and $w$) for $\rho$=0.2 and $\rho$=1.0 .  The variation of 
$\tau^{-1}$ with $F$ (with typical values of $t$, $\lambda$, $w$, $E_s$ \& 
$\rho$) is shown schematically in inset (a) of Fig. 4. Of interest 
is the solution that exists even in the limit $w\lambda^{-1}=0$ (the other 
two solutions do not exist in this limit). The variation of this solution with 
$E_s$ (fixed $t$, $\lambda$ and $w$) for $\rho$=0.2 and $\rho$=1.0 is 
shown schematically in inset (b) of Fig. 4. The two curves show that beyond 
a certain value of $E_s$ a force $-F$ (with $F$ resulting in $\rho$=0.2) 
cannot restore stability. This value is therefore the upper bound ($E_s^u$) 
on $E_s$. Below $E_s^u$ the magnitude of $F$ needed for $\rho$=0.2 is 
larger. The value of $F$ at $E_s^u$ is therefore the lower bound ($F_\ell$) 
on $F$. $F_0=F(E_s=0,\rho=0.2)$.

Fig. 4 shows the variation of $F_\ell$, $E_s^u$ and $F_0$ with the parameters 
$t$ and $\lambda$ ($w=0.15 \mu m$). As noted previously the range of the 
$t$ depends on the value of $\lambda$, with additional restrictions on the 
upper bound of $t$.  These restrictions are due to the absence of the 
solution (of interest) to $\tau^{-1}=0$ for small $\lambda$ and large $t$. This 
seems to correspond the behavior observed at higher 
temperatures \cite{EFu} wherein a current in the step $down$ direction 
causes the step bunching instability.
For $tw\leq 1$, the curves in Fig. 4 correspond to the limit 
$w\lambda^{-1} \rightarrow 0$ and can be obtained by solving $\tau^{-1}=0$ to 
zeroth order in $w\lambda^{-1}$. This gives 

\begin{equation}
F =  \frac{k_bT}{\lambda^2t} \left( \frac{(1+e^{-2E_s/k_bT})-
\rho(1+e^{2E_s/k_bT})}{\rho -1} \right) \
\end{equation} 

\noindent Using the above expression, $E_s^u$, $F_\ell$ and $F_0$ are  
found to be 

\begin{equation}
E_s^u = \frac{k_bT}{2}ln \left( \frac{3+\rho_0}{3\rho_0+1} \right) \
\end{equation}

\begin{equation}
F_\ell = \frac{k_bT}{\lambda^2t} \left( \frac{4(1-\rho_0^2)}{(3+\rho_0)
(3\rho_0+1)} \right) \
\end{equation}

\begin{equation}
F_0=\frac{k_bT}{\lambda^2t} \left( \frac{2(1-\rho_0)}{1+\rho_0} \right) \
\end{equation}

\noindent with $\rho_0=0.2$ (here). Equation 4 gives $E_S^u$=0.05 eV for the 
experiment of Williams {\it et al} \cite{EFu}. This is much smaller 
than the barrier calculated here for the [$\overline{2}11$] step 
(0.61$\pm$0.07 eV) suggesting that [$\overline{11}2$] steps were seen in the 
experiment. From equn. 6 and the electric field in the experiment \cite{EFu} 
(700 V/m) the (maximum) charge needed on the adatoms 
(in units of the (negative) electronic charge) is estimated to be reasonably 
small- in the 
range $3\times10^{-2}$ to $3\times10^{-5.1}$. From Fig. 4 it can be seen that  
with a Schwoebel barrier the charge needed is always smaller than with a zero 
barrier. 
For $tw>1$, the deviation of the curves 
from the above expressions is significant - this is due to the dependence of 
$\tau^{-1}$ on $w\lambda^{-1}$ which require that terms of order 
$w\lambda^{-1}$ and higher be considered. In this range, the minimum charge 
needed on the adatoms (using $F_\ell$) continues to have a reasonable lower 
limit as it is in the range $2\times10^1$ to $7\times10^{-5}$.

\section{Conclusions}

In summary the adatom potential energy contours have been calculated for the 
1$\times$1 reconstructed Si(111) surface and the two high symmetry single 
height steps on it using the empirical Stillinger-Weber potential. 
From these plots diffusion barriers along the path of 
minimum barrier height can be calculated for transitions between any two 
minima. The results show that barriers for  diffusion in the trench along the 
step edge are smaller than that between that global minima on the free surface. 
There is also a strong correlation between that adatom potential energy and the 
potential energy derived from the local geometry of atoms on the adatom free 
surface or 
surface or 
Si(111) surface (0.97$\pm$0.07 eV) and the Schwoebel barrier on the 
[$\overline{2}11$] step (0.61$\pm$0.07 eV) are robust features due to 
changes in adatom coordination number.
Interpreting recent electromigration experiments in the 
presence of a Schwoebel barrier shows that smaller values of electronic charge 
on the adatoms can 
account for the observations.  It also suggests, in the  limit where the 
diffusion 
length ($\lambda$) is much larger than the terrace width ($w$),  
that [$\overline{11}2$] steps 
were observed in the experiment \cite{EFu}. However, this conclusion relies 
on assumptions regarding the nature of the atomic processes at the  
step edge (that these processes may involve two atoms (at a kink site) 
which see 
the barriers as computed here) and that $w\lambda^{-1}\rightarrow 0$. 
Future studies addressing these assumptions may interpret
electromigration experiments differently. 

\bigskip

\section*{Acknowledgments}

One of the authors (S. Kodiyalam) wishes to thank E.D. Williams, N.C. Bartelt,  
Elain Fu, D.J. Liu, D. Kandel and S.V. Khare for useful discussions. This work 
has been supported by the NSF-MRG and the U.S. ONR.
 
\bigskip

\section*{Appendix A: Analysis of the diffusion equation for step motion}
The diffusion equation for non-interacting adatoms subliming in the presence 
of an 
electromigration force $F$ perpendicular to the step edge and the absence of an 
external flux may be written in the ``adiabatic approximation'' (which neglects 
the time derivative of the density) as follows:\cite{Sss,G}

\begin{equation}
\frac{d^2n}{dx^2} - f\frac{dn}{dx} - \frac{n}{\lambda^2} = 0 \ 
\end{equation}

\noindent where $n$ is the density of adatoms, $f=F/k_bT$, $\lambda$ is the 
diffusion length and the positive $x$ axis coincides with the step down 
direction. This equation can be solved using the boundary conditions on the 
current $j$ (=$D \left( -\frac{dn}{dx} + fn \right)$, where $D$ is the 
diffusion constant). These conditions 
determine the step velocity $V$ ( =$a^2(-j_l+j_u)$, where $a^2$ is the inverse 
terrace density, $j_u$ is the current towards the step edge from the upper 
terrace  and $j_l$ is the current away from the step edge on the lower terrace)
and define the step kinetic coefficient $\beta$. In the presence of a 
Schwoebel barrier ($E_s$), this parameter is modeled as assuming the 
values $\beta_0$
and $\beta_1$ at the upper ($x=0$) and lower ($x=W$) terrace step edges 
respectively, with $\frac{\beta_0}{\beta_1}=e^{\epsilon}$, where $\epsilon = 
\frac{2E_s}{k_bT}$. The boundary conditions can now be written as follows: 

\begin{eqnarray}
\left[ -\frac{dn}{dx} + fn \right]_{x=0}  =  -\frac{te^\epsilon}{1+e^\epsilon}
(n(0) - n_e), \; \; \; \; 
\left[ -\frac{dn}{dx} + fn \right]_{x=W}  =   \frac{t}{1+e^\epsilon}
(n(W)-n_e)
\end{eqnarray}

\noindent where $n_e$ is the equilibrium density and $t=\frac{\beta_0+\beta_1}
{a^2Dn_e}$. The solution to the diffusion equation under the above boundary 
conditions is shown below: 

\begin{equation}
n(x) = n_e(Ae^{W\lambda^{-1}\eta_+} + Be^{W\lambda^{-1}\eta_-}) \\
\end{equation}

\noindent with 

\noindent $\eta_\pm=f\lambda/2 \pm \surd(1+(f\lambda/2)^2), 
A =  \frac{\psi e^{W\lambda^{-1}\eta_-}-\varphi}
{\phi\psi e^{W\lambda^{-1}\eta_-}-\varphi\chi e^{W\lambda^{-1}\eta_+}},
B = \frac{\phi-\chi e^{W\lambda^{-1}\eta_+}}
{\phi\psi e^{W\lambda^{-1}\eta_-}-\varphi\chi e^{W\lambda^{-1}\eta_+}} $.

\noindent where

\noindent $\phi = 1 - \frac{(\eta_+ - f\lambda)(1 + e^\epsilon)}
{t\lambda e^\epsilon}$,
$\varphi = 1 - \frac{(\eta_- - f\lambda)(1 + e^\epsilon)}
{t\lambda e^\epsilon}$,
$\chi = 1 + \frac{(\eta_+ - f\lambda)(1 + e^\epsilon)} {t\lambda}$ and 
$\psi = 1 + \frac{(\eta_- - f\lambda)(1 + e^\epsilon)} {t\lambda}$.

\noindent The velocity ($V_n$) of a particular step edge ($X_n$) can now be 
calculated:

$V_n=a^2(-j_l (=j(x=0,W=X_{n+1}-X_n)) + j_u (=j(x=W,W=X_n-X_{n-1}) )$

\noindent From the functional form of the density and using 
$D=\frac{\lambda^2}{\tau_{lf}}$ ($\tau_{lf}$ is the lifetime of an 
adatom) it may be seen that 

\begin{equation}
V_n=\frac{n_ea^2}{\tau_{lf}}\lambda (g_- + g_+) \
\end{equation}
\noindent where $g_\pm$ are dimensionless functions of dimensionless arguments 
given by:

\noindent $g_-(\epsilon,t\lambda,f\lambda,W\lambda^{-1}) = 
t\lambda \left[ \frac{e^\epsilon(A+B-1)}{1+e^\epsilon} \right]$,
$g_+(\epsilon,t\lambda,f\lambda,W\lambda^{-1}) =
t\lambda \left[ \frac{Ae^{W\lambda^{-1}\eta_+}+Be^{W\lambda^{-1}\eta_-}-1}
{1+e^\epsilon} \right]$. 
 
\noindent with the symmetry $g_+(-a,b,-c,d) = g_-(a,b,c,d)$. 

\bigskip

\section*{Appendix B: An atomic model for the step kinetic coefficient} 
The step kinetic coefficient $\beta$ was introduced by Chernov \cite{Ch} 
through the supposition that the currents ($j_l$ from the lower terrace and 
$j_u$ from the upper terrace) towards a step edge (at $x=X$) are proportional 
to the deviation of the adatom density $n$ in the neighborhood of the step 
from the equilibrium density $n_e$. From the model adopted here it will be 
shown that in the presence of a Schwoebel barrier this parameter assumes 
different values {\it i.e.} $\beta_0$ and $\beta_1$ determining $j_l$ and $j_u$ 
respectively: 

\begin{eqnarray}
j_l  =  -\frac{\beta_0}{a^2} \left[ \frac{\theta(X^+) - \theta_e}{\theta_e} 
\right], \; \; \; \;
j_u  =  \frac{\beta_1}{a^2} \left[ \frac{\theta(X^-) - \theta_e}{\theta_e} 
\right] 
\end{eqnarray}

\noindent where $a^2 = 1/$(terrace density),$\theta=na^2$ and $\theta_e=n_ea^2$
with the positive $x$ axis in the step down direction. These equations 
serve as the boundary conditions used in this work (eqn 8).  

The temperature dependence of $\beta_0$ ($\beta_1$) can be derived assuming an 
activated model for adatom attachment-detachment at the step edge from the 
lower (upper) terrace with the 
single atom activation barrier being $E_a$ ($E_a+E_s$) where $E_a$ is the 
barrier on the free surface and $E_s$ is the Schwoebel barrier. \cite{EWp} 
However, since 
silicon has two atoms in its unit cell, $successful$ attachment-detachment 
processes at the kink sites of steps must involve two atoms. Single atom 
processes from energetically favorable step edge configurations 
lead to unfavorable ones and are expected to have very large barriers. Further, 
the probability of simultaneous attachment (detachment) of two atoms is 
proportional to $\theta_e^2$ ($(1-\theta_e)^2$). The currents are therefore 
given by:

\begin{eqnarray}
j_l  =  -\frac{\nu_le^{\frac{-2E_a}{k_bT}}\theta^2 - c_l(1-\theta)^2}
{a_\parallel}, \; \; \; \; 
j_u  =  \frac{\nu_ue^{\frac{-2(E_a+E_s)}{k_bT}}\theta^2 - c_u(1-\theta)^2}
{a_\parallel} \nonumber 
\end{eqnarray}

\noindent where $a_\parallel$ is the lattice constant along the step and 
$\nu_{l,u}$ are ``effective'' attachment attempt frequencies. The vanishing of 
these currents in equilibrium (when $\theta = \theta_e$) determine the 
constants $c_{l,u}$. Further assuming $\nu_l=\nu_u=\nu$ and neglecting terms 
of the order $(\theta - \theta_e)^2$ gives:

\begin{eqnarray}
j_l  =  -\frac{2\nu\theta_ee^{\frac{-2E_a}{k_bT}}}{a_\parallel(1-\theta_e)}
(\theta-\theta_e), \; \; \; \;
j_u  =  \frac{2\nu\theta_ee^{\frac{-2(E_a+E_s)}{k_bT}}}
{a_\parallel(1-\theta_e)}(\theta-\theta_e) \nonumber 
\end{eqnarray}

\noindent from which $\beta_0$ and $\beta_1$ can  be identified. Hence 

\begin{equation}
\beta_0+\beta_1 = \frac{2\nu a^2}{a_\parallel} \left[ 
\frac{ \theta_e^2}{1-\theta_e} \right] (e^{\frac{-2E_a}{k_bT}} + 
e^{\frac{-2(E_a+E_s)}{k_bT}}) 
\end{equation}
\begin{equation}
\frac{\beta_0}{\beta_1} = e^{\frac{2E_s}{k_bT}} 
\end{equation}

\subsection*{Relationship between the step kinetic coefficient and the step 
mobility\cite{NCp}}

To relate the step mobility $\Gamma$ and the step kinetic coefficient 
$\beta_s = \beta_0+\beta_1$,
the expression for the step velocity ($\frac{dX}{dt}=
a^2(j_u - j_l)$) must be obtained in 
terms of $\beta_s$. From Chernov's defining equations (11):

\begin{eqnarray}
\frac{dX}{dt} = \frac{1}{\theta_e}(\beta_1\theta(X^+)+
\beta_0\theta(X_-) - \beta_s\theta_e) 
\end{eqnarray}

\noindent This equation needs to be extended to account for the two dimensional 
character of the step: The currents towards the step edge are proportional to 
the deviation of the chemical potential $\mu$ from the equilibrium chemical 
potential near the step edge $\mu^s_e$. These relationships may be assumed to 
be the defining equations for the step mobilities $\Gamma_1$ and $\Gamma_0$:
\begin{eqnarray}
j_u= \frac{\Gamma_0}{a^4}(\mu(X^-) - \mu^s_e),  \; \; \; \;
j_l= -\frac{\Gamma_1}{a^4}(\mu(X^+) - \mu^s_e)
\end{eqnarray}

\noindent Assuming the existence of a coarse-grained free energy functional 
$\cal F$ of the step configuration $\{x(y)\}$, $\mu^e_s$ is given by 
Mullins \cite{WWM} to be:

\begin{equation}
\mu^s_e=\mu_e + \frac{a^2}{k_bT}\frac{\delta{\cal F}({x(y)})}{\delta x}
\end{equation}
 
\noindent where $\mu_e$ is the equilibrium chemical potential on the terrace 
far away from the step edge. 
Further adopting his model of a dissolving 
substance for the adatoms, and in an approximation in which the deviation of 
the density $\theta$ from the equilibrium density  
$\theta_e$ (the density far away from the step edge) is much smaller than 
$\theta_e$, the chemical potential and the 
density can be related by \cite{WWM}:
\begin{equation} 
\mu(\theta)-\mu_e = k_bT\frac{\theta-\theta_e}{\theta_e} 
\end{equation}
\noindent Using equations 16 and 17 it can be seen that the equilibrium 
density beside the step edge is different from $\theta_e$.\cite{NC2,US,BZ} 
Further using equation 15, the step velocity can be given by: 

\begin{eqnarray}
\frac{\partial x}{\partial t} = \frac{1}{\theta_e}\left(\frac{\Gamma_1}{a^2}
\theta(x^+)+ \frac{\Gamma_0}{a^2}\theta(x^-) - \frac{\Gamma\theta_e}{a^2}
\left(1 + \frac{a^2}{k_bT} \frac{\delta{\cal F}({x(y)})}
{\delta x} \right)\right)
\end{eqnarray}

\noindent where $\Gamma = \Gamma_0 + \Gamma_1$. In the absence of an external 
force on the adatoms and when 
attachment-detachment of atoms from the terrace onto the step edge is rate 
limiting in relation to terrace diffusion it is expected that $\theta(x^+)=
\theta(x^-)=\theta_e$. Then, equation 18 reduces to the Langevin equation 
(without the noise term $\zeta$) \cite{NC1,NC}:

\begin{eqnarray}
\frac{\partial x}{\partial t} = -\frac{\Gamma}{k_bT}
\frac{\delta{\cal F} ({x(y)})}{\delta x} + \zeta (y,t) \nonumber
\end{eqnarray}

\noindent Within the small slope approximation ($\frac{\partial x}{\partial y} 
<< 1$), $\frac{\delta{\cal F}({x(y)})}{\delta x} \propto \frac{\partial^2 x}
{\partial y^2}$.
The spatial average (over y, denoted by $< >$) of this term vanishes due to 
periodic boundary conditions. Identifying $<x>$ with $X$, assuming that the 
density is independent of $y$ and approximating $<\theta(x)> = \theta(X)$,
equation 18 (on spatial averaging) gives:

\begin{equation} 
\frac{dX}{dt}=\frac{1}{\theta_e}\left(\frac{\Gamma_1}{a^2}
\theta(X^+)+ \frac{\Gamma_0}{a^2}\theta(X^-) - \frac{\Gamma\theta_e}{a^2}
\right)
\end{equation}

\noindent Equations 14 and 19 agree if 

\begin{eqnarray}
\beta_{0,1}=\frac{\Gamma_{0,1}}{a^2} \; \; \Rightarrow \; \; \beta_s=
\frac{\Gamma}{a^2} 
\end{eqnarray}

\clearpage

\begin{figure}
\caption{One bilayer of the Si(111) surface consisting of the upper monolayer 
(grey) and lower monolayer (black). The figure shows the threefold and 
reflection symmetry of this surface: Steps running along directions with 
equal $\theta$ are identical. (Borrowed from reference [21])}
\end{figure}
\begin{figure}
\caption{The Si(111) surface (a) (the upper monolayer is shown in grey and the 
lower monolayer in black), the [$\overline{2}11$] step (d) and the 
[$\overline{1}\overline{1}2$] step (g) (for the step configurations the upper 
bilayer is shown with larger atoms as compared to the lower bilayer).  
(b), (e) \& (h) show the corresponding
adatom binding energy $V(x,y)$. (c), (f) \& (i) show the corresponding 
binding energy derived from the local geometry $V_{lg}(x,y)$. Minima, 
saddle points and maxima are marked (labeled) by +(m), *(s) and $\times$(M) 
respectively with the figure in parenthesis being their corresponding 
value in eV.
In (b) \& (c) 
the contours are 0.1 eV apart. In (e), (f), (h) \& (i) they are 0.2 eV 
apart. In (e) \& (h) contours of $V \geq$-2.2 eV and in (f) \& (i) those 
of $V_{lg} \geq$-1.9 eV are marked with dashed lines. The contour plots 
suggest a strong correlation between $V$ and $V_{lg}$. The  diffusion 
barrier on the surface is determined by m1 \& s1 in (b) \& (c). The 
Schwoebel barrier is determined by s3 in (e) \& (f) whereas it is determined 
by s5 in (h) and s6 in (i). There is a 
one-to-one correspondence between the barrier determining features in $V(x,y)$ 
and  $V_{lg}(x,y)$ for the Si(111) surface and the [$\overline{2}11$] step 
suggesting that the Schwoebel barrier on the [$\overline{2}11$] step is 
robust feature. }
\end{figure}
\begin{figure}
\caption{Plots of the adatom potential energy derived from the local
geometry of fixed atoms $V_{lg}$ $vs$ the actual adatom potential energy $V$.
The straight line fits show the strong correlation between $V_{lg}$ and $V$.
The rough configuration independence of the relationship suggests that it is
a characteristic of the (111) surface.}
\end{figure}
\begin{figure}
\caption{Inset (a) shows schematically the variation of $\tau^{-1}=k_+-\rho k_-$
with $F$. The solution to $\tau^{-1}=0$ 
(of interest) is $F_r$ which exists even in the limit $w\lambda^{-1}=0$ 
whereas the other two do not. Inset (b) shows schematically the variation of 
$F_r$ with $E_s$ for $\rho=1.0$ and $\rho=0.2$. The upper bound on $E_s$ 
($E_s^u$) is determined from the condition $F(E_s > E_s^u,\rho=1.0) < 
-F(E_s \geq E_s^u,\rho=0.2)$. It can be seen that $F(E_s^u,\rho=0.2)$ is the 
lower bound on $F$ ($F_\ell$). $F_0=F(E_s=0,\rho=0.2)$. For the curves shown, 
$w=0.15\mu m$, $0.5\mu m \leq \lambda \leq0.5\times10^3\mu m$ 
(7 values equally spaced on the logarithmic scale) with the range of $t$ 
depending on $\lambda$: $\frac{7.7\times 10^3}{\lambda^2}\leq t\leq
\frac{7.7\times10^3}{\lambda^2}\times 10^{3.1}$ ($\lambda>7\mu m$). The upper 
(lower) limits of $t$ are marked by + ($\times$). For $\lambda<7\mu m$ the 
upper limit of $t$ is forced to be smaller than that estimated since the 
solution $F_r$ ceases to exist. For $\lambda<1\mu m$, $F_r$ does not exist 
for even the smallest $t$. This shows that $\lambda_{min}=1 \mu m$. For 
$w\lambda^{-1}\rightarrow 0$, $E_s^u=0.35k_bT$=0.05 eV (for the experiment in
ref. 5) - significantly lesser 
than that calculated here for the [$\overline{2}11$] step (0.61$\pm$0.07 eV). 
This suggests that [$\overline{11}2$] steps were observed. } 
  
\end{figure}

\end{document}